\documentclass[sigconf]{acmart}

\usepackage{pifont}
\newcommand{\cmark}{\ding{51}}%
\newcommand{\xmark}{\ding{55}}%

\usepackage{cleveref}
\usepackage{soul}

\AtBeginDocument{%
  \providecommand\BibTeX{{%
    \normalfont B\kern-0.5em{\scshape i\kern-0.25em b}\kern-0.8em\TeX}}}

\copyrightyear{2024}
\acmYear{2024}
\setcopyright{acmlicensed}\acmConference[CHI '24]{Proceedings of the CHI Conference on Human Factors in Computing Systems}{May 11--16, 2024}{Honolulu, HI, USA}
\acmBooktitle{Proceedings of the CHI Conference on Human Factors in Computing Systems (CHI '24), May 11--16, 2024, Honolulu, HI, USA}
\acmDOI{10.1145/3613904.3642621}
\acmISBN{979-8-4007-0330-0/24/05}

\begin{document}

\setlength{\fboxsep}{8pt}

\title{Explanations, Fairness, and Appropriate Reliance in Human-AI Decision-Making}

\author{Jakob Schoeffer}
\email{schoeffer@utexas.edu}
\orcid{0000-0003-3705-7126}
\affiliation{
    \institution{University of Texas at Austin}
    \city{Austin}
    \country{USA}
}

\author{Maria De-Arteaga}
\authornote{Both authors contributed equally to this research.}
\email{dearteaga@utexas.edu}
\orcid{0000-0003-2297-3308}
\affiliation{%
    \institution{University of Texas at Austin}
    \city{Austin}
    \country{USA}
    }

\author{Niklas Kühl}
\authornotemark[1]
\email{kuehl@uni-bayreuth.de}
\orcid{0000-0001-6750-0876}
\affiliation{%
    \institution{University of Bayreuth}
    \institution{Fraunhofer FIT}
    \city{Bayreuth}
    \country{Germany}
    }

\begin{abstract}
    In this work, we study the effects of feature-based explanations on distributive fairness of AI-assisted decisions, specifically focusing on the task of predicting occupations from short textual bios. We also investigate how any effects are mediated by humans' fairness perceptions and their reliance on AI recommendations. Our findings show that explanations influence fairness perceptions, which, in turn, relate to humans' tendency to adhere to AI recommendations. However, we see that such explanations do not enable humans to discern correct and incorrect AI recommendations. Instead, we show that they may affect reliance irrespective of the correctness of AI recommendations. Depending on which features an explanation highlights, this can foster or hinder distributive fairness: when explanations highlight features that are task-irrelevant and evidently associated with the sensitive attribute, this prompts overrides that \textit{counter} AI recommendations that align with gender stereotypes. Meanwhile, if explanations appear task-relevant, this induces reliance behavior that \textit{reinforces} stereotype-aligned errors. These results imply that feature-based explanations are not a reliable mechanism to improve distributive fairness.
\end{abstract}



\maketitle

\section{Introduction}
AI systems are commonly used for assisting decision-making in consequential areas, where they provide human decision-makers with decision recommendations.
The human is then tasked to decide whether to adhere to such recommendations or override them.
Researchers, policy makers, and activists have expressed concern over the risk of algorithmic bias resulting in unfair decisions.
As a response, many have advocated for the need for explanations, under the assumption that they can enable humans to mitigate algorithmic bias.
For instance, in a recent Forbes article~\citep{kitepowell2022trending}, it is claimed that ``companies [in financial services and insurance] are using explainable AI to make sure they are making fair decisions about loan rates and premiums.''
Others have claimed that explanations ``provide a more effective interface for the human-in-the-loop, enabling people to identify and address fairness and other issues''~\citep{dodge2019explaining}.
However, there is no reliable empirical evidence as to whether existing explainability techniques can live up to these hopes~\citep{langer2021we,deck2023critical}.

Previous research on AI-assisted decision-making has studied how explanations affect people's fairness perceptions and trust (see, e.g.,~\citep{starke2021fairness,vereschak2021evaluate}).
Prior work has also studied how explanations affect the accuracy of AI-assisted decision-making (e.g., \citep{lai2019human,alufaisan2021does,zhang2020effect}), and people's overall reliance\footnote{We use \emph{reliance} as an umbrella term for people's behavior of adhering to or overriding AI recommendations~\citep{lai2021towards}.} behavior (e.g., \citep{bansal2021does,bussone2015role,poursabzi2021manipulating}).
However, the effect of explanations on distributive fairness of AI-assisted decision-making, and the mechanisms underlying such an effect, have not been studied.
Consider the example of financial services referenced above.
Grounded on the algorithmic fairness literature, one can argue that whether decisions to allocate loans are ``fair'' refers to the \emph{distributive fairness} properties of those decisions~\citep{saxena2020fairness}.
For instance, if there is a demographic group that is systematically denied loans because they are incorrectly predicted to be likely to default on a loan, that could be considered unfair because the decisions incur a high rate of false negatives for that group~\citep{hardt2016equality,liu2018delayed}.
Thus, a human-in-the-loop that addresses such fairness issues should have the capacity to identify mistaken recommendations, reducing the false negative errors affecting that group.
In this case, the goal of explanations should be to help humans identify such errors, yielding AI-assisted decisions that have better distributive fairness properties than the AI alone.
Note that this is different to the \emph{perceptions} that humans may have of an AI system, and it also differs from the overall accuracy or reliance behavior.
In this work, we directly assess how feature-based explanations affect distributive fairness, and the mechanisms underlying this effect.
We empirically study this via a randomized online experiment using a popular task in algorithmic fairness and human-AI interaction studies: predicting a person's occupation from short textual bios.

\paragraph{Our work}
We conduct a first comprehensive analysis of the effects of feature-based explanations on people's ability to enhance distributive fairness in AI-assisted decision-making---and how these effects are mediated by fairness perceptions and reliance on AI recommendations.
To empirically study this, we conduct a randomized between-subjects online experiment and assess differences in perceptions and reliance behavior when participants see and do not see explanations, and when these explanations indicate the use of sensitive features in predictions vs. when they indicate the use of task-relevant features.
We operationalize this by training two AI models with access to different vocabularies.
We conduct our study in the context of occupation prediction based on short textual bios, which is an important task in AI-assisted hiring but at the same time susceptible to gender bias and discrimination~\citep{de2019bias,bogen2018help,sanchez2020does}.
We randomly assign participants to one of two groups and ask them to predict whether bios belong to professors or teachers: for one group, recommendations come from an AI model that uses \emph{gendered} words for predicting occupations, whereas in the other group the AI model uses \emph{task-relevant} words.
Both AI models provide the same recommendations, and their distribution of errors is in line with societal stereotypes and the expected risks of bias characterized in previous research~\cite{de2019bias}.
Participants in both conditions are provided with explanations that visually highlight the most predictive words of their respective AI models.
We also include a baseline condition where no explanations are shown.
We test for differences in perceptions and reliance behavior across conditions, and measure gender disparities for different types of errors.

\paragraph{Findings and implications}
\textbf{First}, we do not observe any significant differences in decision-making accuracy across conditions, i.e., participants did not make more (or less) accurate decisions in the conditions with explanations compared to the baseline without explanations.
Since participants were incentivized to make accurate predictions, this implies that explanations did not enable them to make better decisions with respect to accuracy.

\textbf{Second}, no condition improved participants' likelihood to override mistaken vs. correct AI recommendations, but conditions did affect the likelihood to override AI recommendations conditioned on the predicted occupation: we see that participants in the \emph{gendered} condition overrode more AI recommendations to \emph{counter} existing societal stereotypes (e.g., by predicting more women to be professors), irrespective of whether the prediction was correct.
Simultaneously, when explanations highlight only task-relevant words, reliance behavior \emph{reinforced} stereotype-aligned decisions; e.g., by predicting more men to be professors, even when they are teachers.

This, \textbf{third,} has implications for distributive fairness: by prompting reliance behavior that either counters or reinforces societal stereotypes embedded in AI recommendations, $(i)$ explanations that highlight gendered words led to a \emph{decrease} in error rate disparities (i.e., fostering distributive fairness), whereas $(ii)$ explanations that highlight task-relevant words led to an \emph{increase} in error rate disparities (i.e., hindering distributive fairness).
These findings emphasize the need to differentiate between improved distributive fairness that is driven by a shift in the types of errors vs. improvements that are driven by humans' ability to override mistaken AI recommendations. 

\textbf{Fourth}, we confirm prior works' findings by observing that people's fairness perceptions are significantly lower when explanations highlight gendered words compared to task-relevant words, and empirically show that people override significantly more AI recommendations when their fairness perceptions are low.
However, we observe that perceptions solely relate to the quantity of overrides and do \emph{not} correlate with an ability to discern correct and incorrect AI recommendations.
Hence, fairness perceptions are only a meaningful proxy for distributive fairness when it is desirable to override the AI based on its use of sensitive features.
However, prior research has shown that the idea of ``fairness through unawareness,'' which deems an AI system to be fair if it does not make use
of information that is evidently indicative of a person’s demographics, is neither a necessary nor sufficient condition for distributive fairness~\citep{apfelbaum2010blind,kleinberg2018algorithmic,dwork2012fairness,pedreshi2008discrimination,corbett2018measure,nyarko2021breaking}.

\paragraph{Recommendations for researchers and practitioners}
Overall, our work has direct implications for researchers and designers of socio-technical systems for decision support.
First, we suggest to measure the effects of decision support interventions, such as explanations, on human reliance behavior with respect to meaningful endpoints.
Importantly, when concerned with fairness, we show that humans may foster or hinder distributive fairness even in the \textit{absence} of an ability to override mistaken AI recommendations, by merely shifting errors.
%
%
%
In that regard, our study design serves as a valuable blueprint for assessing the suitability of other types of interventions as potential pathways towards improved distributive fairness of AI-assisted decisions.
We not only emphasize the importance of grounding the design of explanations in specific desiderata, such as distributive fairness, but also of designing explanations in a way that they provide relevant cues which empower humans to achieve these goals.
With respect to fostering distributive fairness, our work suggests that explanations must transcend a mere human-in-the-loop operationalization of the idea of ``fairness through unawareness'' and provide insights to humans that go beyond the fact of whether or not an AI system makes use of sensitive information.
To that point, our findings raise doubts about the reliability of popular feature-based explanations as enablers for distributive fairness.
Instead, we advocate for a more holistic perspective on algorithmic transparency, one that is grounded in concrete desiderata of relevant human stakeholders and encourages thoughtful design choices to enable truly effective human-AI collaboration.
%

\section{Background}\label{sec:background}
In this section, we provide relevant background on our work and review related literature on explanations, reliance, and fairness.
In \Cref{sec:rw_explanations}, we first show how explanations have been touted for their importance towards enhancing different desiderata in AI-assisted decision-making, including fairness.
We then briefly review the types of explanations that are relevant to our work, and we summarize some of the existing critique of explanations that our work adds to.
In \Cref{sec:explanations_reliance}, we revisit prior work on the effects of explanations on human reliance behavior. Here, we see that previous findings have been mostly inconclusive, and that research on explanations' effects on distributive fairness is lacking. Finally, we discuss how prior work has sometimes conflated human attitudes with behavior, and why that is problematic.
Regarding fairness, in \Cref{sec:explanations_fairness}, we show that there is a gap between common claims that commend explanations as an enabler for distributive fairness and the fact that prior work has primarily studied effects on fairness \textit{perceptions}---a gap we are filling with this work. Lastly, we summarize relevant known relationships between fairness perceptions and the use of sensitive features by AI systems, which inform our study design.

\subsection{Explanations of AI}\label{sec:rw_explanations}

\paragraph{Goals of explanations}
AI systems are becoming increasingly complex and opaque, and researchers and policymakers have called for explanations to make AI systems more understandable to humans~\citep{miller2000women,langer2021we,GDPR}.
Apart from the central aim of facilitating human understanding, prior research has formulated a wealth of different desiderata that explanations are to provide, most of which center one or more different types of stakeholders of AI systems~\citep{langer2021we,preece2018stakeholders,ehsan2020human}.
For instance, system designers might be interested in facilitating trust in their systems through explanations, whereas a regulator likely wants to assess a system's compliance with moral and ethical standards~\citep{langer2021we}.
Different goals may sometimes be impossible to accomplish simultaneously~\citep{springer2019making}.
For a comprehensive overview of different aims of explanations, we refer the reader to~\citet{langer2021we} and~\citet{lipton2018mythos}.
Relevant to our work are several desiderata that concern explanations as an alleged means for better and fairer AI-assisted decision-making~\citep{adadi2018peeking,dodge2019explaining}. Importantly, many of such claims are lacking nuance~\citep{deck2023critical}, which motivates our work. For instance, prior work has claimed that ``explainability can be considered as the capacity to reach and guarantee fairness in ML models''~\citep{arrieta2020explainable}. Yet, both explainability and fairness are multi-dimensional concepts, and it is often unclear what it means to improve fairness through explanations, as well as a lack of evidence studying whether this is possible. In this work, we empirically study whether feature-based explanations can enable humans with discretionary power to improve relevant distributive fairness properties of AI-assisted decisions.

\paragraph{Types of explanations}
The scientific literature distinguishes explanations that aim at explaining individual predictions (\emph{local} explanations) from those that aim at explaining the general functioning of an AI model (\emph{global} explanations)~\citep{guidotti2018survey}.
However, it has been argued that combining local explanations can also lead to an understanding of global model behavior~\citep{lundberg2020local}.
So-called \emph{local model-agnostic} explanations, such as LIME~\citep{ribeiro2016should} or SHAP~\citep{lundberg2017unified}, have gained popularity in the literature~\citep{adadi2018peeking}.
When applied to text data, these methods can generate a highlighting of important words for text classification.
In this work, our focus is on these feature-based explanations, and we use LIME in our experiments, due to its popularity in the literature as well as in practice~\citep{elshawi2021interpretability,bhatt2020explainable,gilpin2018explaining} and, importantly, the fact that LIME has been claimed to enable fairness assessments (e.g., \citep{alves2021making,chakraborty2020making,bhargava2020limeout}), which we challenge in this work.

\paragraph{Criticism of explanations}
Most desiderata for explanations are insufficiently studied or met with inconclusive or seemingly contradictory empirical findings~\citep{langer2021we,chen2023understanding,de2022perils}.
A major line of criticism stems from the fact that explanations can mislead people: \citet{chromik2019dark} discuss situations where system designers may create interfaces or misleading explanations to purposefully deceive more vulnerable stakeholders like auditors or decision-subjects; e.g., through \emph{adversarial attacks} on explanation methods~\citep{slack2020fooling,lakkaraju2020fool,pruthi2019learning,dimanov2020you}.
In the extreme case of placebic explanations (i.e., explanations that convey no information about the underlying AI), \citet{eiband2019impact} find that people may exhibit levels of trust similar to ``real explanations.''
This shows that the sheer presence of explanations can increase people's trust in AI.
With respect to reliance behavior, \citet{banovic2023being} show how explanations can be exploited to exaggerate an AI system's capabilities and, as a result, make humans rely on its recommendations more.
Even in the absence of any malicious intents,~\citet{ehsan2021explainability} highlight several challenges arising from unanticipated negative downstream effects of explanations, such as misplaced trust in AI, or over- or underestimating the AI's capabilities.
In the context of fairness, feature-based explanations may or may not highlight the usage of sensitive information (e.g., on gender) by an AI system, which has been shown to be an unreliable indicator of a system's actual fairness~\citep{apfelbaum2010blind,pedreshi2008discrimination,corbett2018measure,dwork2012fairness,kleinberg2018algorithmic,nyarko2021breaking}.
We address this in more detail in \Cref{sec:explanations_fairness} due to its importance for our work.
Overall, explanations have been shown to be unreliable mechanisms with respect to fostering desiderata like calibrated trust. In this work, we study whether or not this is also the case for distributive fairness.

\subsection{Explanations and (Appropriate) Reliance}\label{sec:explanations_reliance}

\paragraph{Effects on accuracy}
It has been argued that explanations are an enabler for better AI-assisted decision-making~\citep{arrieta2020explainable,dodge2019explaining,kizilcec2016much,gilpin2018explaining,rader2018explanations}.
A recent meta-study~\citep{schemmer2022meta} on the effectiveness of explanations, however, implies that explanations in most empirical studies did not yield any significant benefits with respect to decision-making accuracy; e.g., in~\citep{alufaisan2021does,green2019principles,narayanan2018humans,liu2021understanding,zhang2020effect}.
On the other hand, \citet{lai2019human} find that explanations greatly enhance decision-making accuracy for the case of deception detection.
An accuracy increase through explanations may, however, solely be due to $(i)$ an overall increase in adherence to a high-accuracy AI, or $(ii)$ an overall decrease in adherence to a low-accuracy AI~\citep{bansal2021does,schoeffer2023interdependence}. Importantly, even if explanations lead to more (in)accurate decisions, it is unclear from prior work how this relates to distributive fairness properties. In this paper, we show that changes to distributive fairness metrics may occur even in the absence of any effects on accuracy.

\paragraph{Effects on reliance}
In the context of AI-assisted decision-making, \emph{appropriate reliance} is typically understood as the behavior of humans of overriding incorrect AI recommendations and adhering to correct ones~\citep{schoeffer2023interdependence,schemmer2023appropriate}.
Humans' ability to override mistaken recommendations has also been referred to as \emph{corrective overriding}~\citep{de2020case}.
When considering the role of explanations in fostering appropriate reliance, it has been claimed that ``transparency mechanisms also function to help users learn about how the system works, so they can evaluate the \emph{correctness} of the outputs they experience and identify outputs that are incorrect''~\citep{rader2018explanations}.
Empirical evidence, however, is less clear: several studies have found that explanations can be detrimental to appropriate reliance~\citep{poursabzi2021manipulating,bansal2021does,bussone2015role,schemmer2022influence,van2021evaluating,lai2021towards}, when they increase or decrease humans' adherence to AI recommendations regardless of their correctness.
These phenomena are commonly referred to as \emph{over-} or \emph{under-reliance}~\citep{schoeffer2023interdependence}.
Our work also studies people's reliance behavior, but with a focus on how it relates to distributive fairness---which has not been studied before. In that regard, we show that explanations may foster reliance behavior that either reinforces or counters stereotypical AI recommendations, independent of the correctness of said recommendations.

\paragraph{Conflation of reliance and trust}
Our work is also motivated by the fact that prior work has often conflated human attitudes and behavior.
For instance, many studies have treated reliance and trust interchangeably~\citep{lai2021towards}, sometimes calling reliance a ``behavioral trust measure''~\citep{papenmeier2022s}.
However, definitions of \emph{trust} are often inconsistent~\citep{papenmeier2022s,lee2004trust,jacovi2021formalizing}, which makes empirical findings challenging to compare.
More importantly, trust and reliance are different constructs~\citep{lai2021towards}: reliance is the \emph{behavior} of adhering to or overriding AI recommendations, whereas trust is a subjective \emph{attitude} regarding the whole system, which builds up and develops over time~\citep{rempel1985trust,yu2017user,parasuraman1997humans}.
It has been argued that trust may impact reliance~\citep{dzindolet2003role,lee2004trust,shin2019role}, but trust is not a sufficient requirement for reliance when other factors, such as time constraints, perceived risk, or self-confidence, impact decision-making~\citep{lee2004trust,riley2018operator,de2020case}.
In our work, we directly measure participants' reliance behavior and do not assume an equivalence between reliance and trust.

\subsection{Explanations and Fairness}\label{sec:explanations_fairness}

\paragraph{Goal of promoting algorithmic fairness}
It is known that AI systems can issue predictions that may result in disparate outcomes or other forms of injustices for certain socio-demographic groups---especially those that have been historically marginalized~\citep{de2022algorithmic,imana2021auditing,buyl2022tackling,bartlett2022consumer}.
When AI systems are used to inform consequential decisions, it is important that a human can override problematic recommendations.
To that end, the literature has often framed explanations as an important pathway towards improving algorithmic fairness~\citep{langer2021we,arrieta2020explainable,dodge2019explaining,das2020opportunities}.
Grounded on the organizational justice literature~\citep{colquitt2015measuring,greenberg1987taxonomy}, researchers distinguish different notions of algorithmic fairness, among which are $(i)$ \emph{distributive fairness}, which refers to the fairness of decision outcomes~\citep{zafar2017fairness}, and $(ii)$ \emph{procedural fairness}, which refers to the fairness of decision-making procedures~\citep{lee2019procedural}.
Distributive fairness is typically measured in terms of statistical metrics such as parity in error rates across groups~\citep{barocas-hardt-narayanan,chouldechova2017fair}; which is closely related to notions like \emph{equalized odds} or \emph{equal opportunity}~\citep{hardt2016equality}.
In this work, we apply this notion and measure distributive fairness as disparities in error rates across genders (see \Cref{subsec:measuring_reliance_fairness}).
Importantly, there is no conclusive evidence from prior work showing that explanations lead to fairer decisions, and it remains unclear \emph{how} explanations may enable this~\citep{langer2021we}.
To understand these mechanisms better, we therefore conduct a comprehensive analysis of explanations' effects on the human ability to improve distributive fairness, and we also study in depth the mediating roles of fairness perceptions and reliance behavior.

\paragraph{Fairness perceptions}
Prior work at the intersection of fairness and explanations has primarily focused on assessing how people \emph{perceive} the fairness of AI systems~\citep{starke2021fairness,lai2021towards}.
Empirical findings are mostly inconclusive, stressing that fairness perceptions depend on many factors, such as the explanation style~\citep{binns2018s,dodge2019explaining}, the amount of information provided~\citep{schoeffer2022there}, the use case~\citep{angerschmid2022fairness}, user profiles~\citep{dodge2019explaining}, the decision outcome~\citep{shulner2022fairness}, or whether the final decision is made by a human or an algorithm~\citep{mok2023people}.
Surprisingly, few works have examined downstream effects of fairness perceptions on AI-assisted decisions, which has also been noted by \citet{starke2021fairness}.
Our work complements prior studies by centering distributive fairness and how it relates to fairness perceptions.
Importantly, in this work we stress the importance of not conflating fairness perceptions with the ability to improve distributive fairness.

\paragraph{Perceptions and sensitive features}
A series of prior studies have found that knowledge about the features that an AI model uses influences people's fairness perceptions~\citep{grgic2016case,grgic2018beyond,grgic2018human,van2019crowdsourcing,plane2017exploring,nyarko2021breaking}.
This type of information is, e.g., conveyed by feature-based explanations like LIME.
Specifically, people tend to be averse to the use of what is typically considered \emph{sensitive} information, e.g., gender or race~\citep{grgic2018human,grgic2018beyond,grgic2016case,corbett2018measure,plane2017exploring,schoeffer2022there,nyarko2021breaking}.
Interestingly, people's perceptions towards these features may change after they learn that ``blinding'' the AI model to these features can lead to \emph{worse} outcomes for marginalized groups~\citep{nyarko2021breaking}.
Similarly, it has been shown that people's perceptions towards the inclusion of sensitive features switch when they are told that this inclusion makes an AI model more accurate~\citep{grgic2016case} or equalizes error rates across demographic groups~\citep{harrison2020empirical}.
In fact, it is known that prohibiting an AI model from using sensitive information is neither a necessary nor sufficient requirement for fair decision-making~\citep{apfelbaum2010blind,kleinberg2018algorithmic,dwork2012fairness,pedreshi2008discrimination,corbett2018measure,nyarko2021breaking}, and that there exist several real-world examples where the inclusion of sensitive features can make historically disadvantaged groups like Black people or women better off~\citep{pierson2020large,corbett2018measure,skeem2016gender,mayson2018bias}.
In this work, we build upon these findings on the interplay of fairness perceptions and sensitive features.
Concretely, we assess differences in reliance behavior when participants see explanations that highlight task-relevant vs. sensitive features, and derive implications for distributive fairness.

\section{Study design}\label{sec:experimental_design}
In this section, we outline our study design. First, we introduce the task and dataset for our study, then we explain the experimental setup and our dependent variables, and, finally, the data collection process.

\subsection{Task and Dataset}\label{sec:task_dataset}

\begin{figure*}[t]
\centering
\begin{minipage}{0.48\textwidth}
    \centering
    \includegraphics[width=0.95\linewidth]{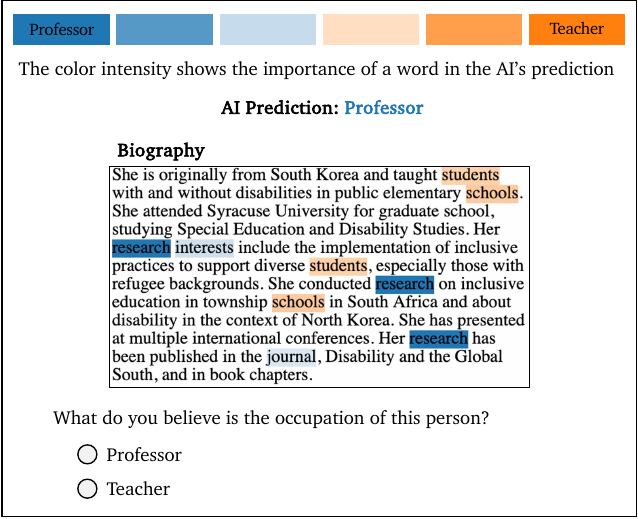}
    \caption*{\emph{Task-relevant} condition}
\end{minipage}
\begin{minipage}{0.48\textwidth}
    \centering
    \includegraphics[width=0.95\linewidth]{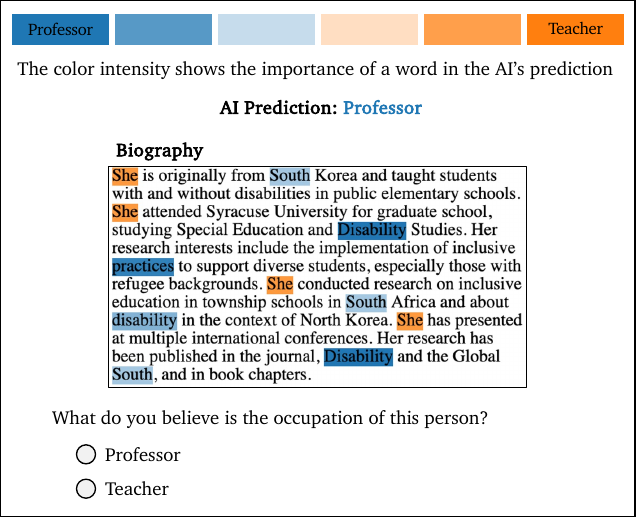}
    \caption*{\emph{Gendered} condition}
\end{minipage}
\caption{A bio of a woman professor, both in the \emph{task-relevant} (left) and the \emph{gendered} (right) condition.}
\label{fig:exemplary_conditions}
\Description{A bio of a woman professor in both conditions.}
\end{figure*}

\paragraph{Task}
Automating parts of the hiring funnel has become common practice of many companies; especially the sourcing of candidates online~\citep{bogen2018help,sanchez2020does}.
An important task herein is to determine someone's occupation, which is a prerequisite for advertising job openings or recruiting people for adequate positions.
This information may not be readily available in structured format and would, instead, have to be inferred from unstructured information found online.
While this process lends itself to the use AI systems, it is susceptible to gender bias and discrimination~\citep{de2019bias,bogen2018help,sanchez2020does}.
\citet{de2019bias} show that these biases can manifest themselves in error rate disparities between genders, and that error rate disparities are correlated with gender imbalances in occupations.
For instance, women surgeons are significantly more often misclassified than men surgeons because the occupation \emph{surgeon} is heavily men-dominated.
Similar disparities occur, among others, for professors and teachers.
Interestingly, the disparate impact on people persists when the AI model does \emph{not} consider explicit gender indicators (e.g., pronouns)~\citep{de2019bias}.
Such misclassifications in hiring have tremendous repercussions for affected people because they may be systematically excluded from exposure to relevant opportunities.
In our study, we instantiate an AI-assisted decision-making setup where participants see short textual bios and are asked---with the help of an AI recommendation---to predict whether a given bio belongs to a professor or a teacher.
Professors are historically a men-dominated occupation, whereas teachers have been mostly associated with women~\citep{miller2000women}.\footnote{See also~\citep{teachers,professors} on current demographic statistics for professors and teachers in the US.}

\paragraph{Dataset}
We use the publicly available BIOS dataset, which contains approximately 400,000 online bios for 28 different occupations from the Common Crawl corpus, initially created by~\citet{de2019bias}.\footnote{The code that reproduces the dataset can be found at \url{https://github.com/Microsoft/biosbias}.}
This data set has been used in other human-AI decision-making studies as well, such as the ones by~\citet{liu2021understanding} or~\citet{peng2022investigations}.
For each bio in the dataset we know the gender of the corresponding person and their true occupation.
Gender is based on the pronouns used in the bio, and a limitation of this dataset is that it only contains bios that use ``she'' or ``he'' as pronouns, excluding bios of non-binary people.
We only consider bios that belong to professors and teachers, which leaves us with 134,436 bios, out of which 118,215 belong to professors and 16,221 to teachers.
In line with current demographics and societal stereotypes~\citep{teachers,professors,miller2000women}, we have more men (55\%) than women (45\%) bios of professors and more women (60\%) than men (40\%) bios of teachers.

\subsection{Experimental Setup}\label{sec:study_setup}

\paragraph{General procedure}
We conduct a between-subjects study where participants see 14 bios one by one, each including the AI recommendation as well as an explanation highlighting the most predictive words.
We also include a baseline condition without explanations.
The crux of our experimental design is that we assign participants to conditions where they see recommendations and explanations either from $(i)$ an AI model that uses \emph{task-relevant} features, or $(ii)$ an AI model that uses \emph{gendered} (i.e., sensitive) features.
An exemplary bio including explanations is depicted in \Cref{fig:exemplary_conditions}.
Note that the AI predictions and explanations stem from actual AI models that agree in their predictions for the 14 bios shown to participants; we outline the construction of these models later in this section as well as, more extensively, in \Cref{app:construction_tr_g}.

\begin{figure*}[t]
    \centering
    \includegraphics[width=0.55\textwidth]{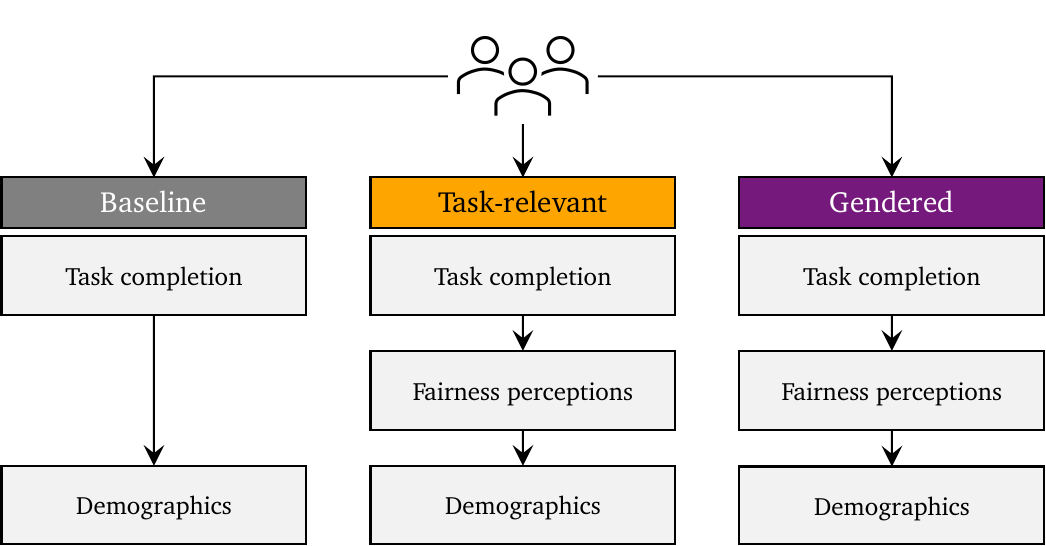}
    \caption{Study participants are randomly assigned to one of three conditions. In each condition, they first complete the task of predicting occupations from 14 short bios, and complete a demographic survey. In the conditions with explanations (\textit{Task-relevant} and \textit{Gendered}), participants are also asked about their fairness perceptions after completing the task.}
    \label{fig:study_setup}
    \Description{Study participants are randomly assigned to one of three conditions. In each condition, they first complete the task of predicting occupations from 14 short bios, and complete a demographic survey. In the conditions with explanations, participants are also asked about their fairness perceptions after completing the task.}
\end{figure*}

Participants in each condition first complete the task of predicting occupations for 14 bios, and---if assigned to a condition with explanations---answer several questions regarding their fairness perceptions.
Since the baseline condition does not provide any cues regarding the AI's decision-making procedures, we do not ask about perceptions there.
Finally, participants provide some demographic information.
A summary of our general setup in illustrated in \Cref{fig:study_setup}.
Note that we ask about fairness perceptions \emph{after} the task is completed, so as to prevent these questions from moderating reliance behavior~\citep{chaudoin2021survey}.
Given that distinguishing professors and teachers based on their bios can be at times ambiguous and not everyone may be familiar with the differences, we also ask at the beginning of our questionnaires what participants consider the difference between \emph{professor} and \emph{teacher} to be.
Additionally, after completing the task, we ask participants an open-ended question on what information they relied on when differentiating \textit{professor} and \textit{teacher}.
This way, we were able to confirm---both quantitatively and qualitatively---that participants thought consistently about this distinction between conditions.

\paragraph{Task completion}
\Cref{fig:exemplary_conditions} shows the interface that participants in the \emph{task-relevant} as well as the \emph{gendered} condition see during the completion of the task.
In line with traditional LIME applications for text data~\citep{mardaoui2021analysis}, explanations involve a dynamic highlighting of important words for either AI model (\emph{task-relevant} and \emph{gendered}).
For a given prediction, the colors indicate which outcome prediction (\textit{professor} or \textit{teacher}) a word contributes towards (see~\citep{ribeiro2016should}). The binary prediction is then obtained using a threshold applied to the predicted probability.
For instance, in \Cref{fig:exemplary_conditions} (\textit{task-relevant} condition), the word \textit{students} is indicative of \textit{teacher} (orange), and the word \textit{research} is indicative of \textit{professor} (blue). The binary prediction is \textit{professor} because the predicted probability that the bio belongs to a professor is greater to 50\%.
Lastly, the color intensity shows the importance of a given word in the AI's prediction.
This interface is similar to related studies on AI-assisted text classification~\citep{lai2020chicago,schemmer2023appropriate,liu2021understanding}.
Participants in the \emph{task-relevant} and the \emph{gendered} condition are confronted with 14 bios similar to the one in \Cref{fig:exemplary_conditions}, whereas participants in the baseline condition are shown the same set of bios without highlighting of words, and the AI prediction without color coding.
Recall that the AI recommendations are identical across conditions.
For each instance, participants are asked to make a binary prediction about whether they believe that a given bio belongs to a professor or a teacher.
We incentivize accurate predictions through bonus payments (see \Cref{sec:data_collection}).

\paragraph{Task-relevant and gendered classifiers}
We provide intuition for how we constructed the AI models that generate recommendations and explanations in the \emph{task-relevant} and \emph{gendered} conditions.
We defer a detailed explanation to \Cref{app:construction_tr_g}.
The general idea is to train two classifiers with access to mutually disjoint vocabularies as predictors.
The \emph{task-relevant} vocabulary consists of words that appear on average---for both men and women---more often in professor or teacher bios than in any of the 26 remaining occupations in the BIOS dataset.
The resulting vocabulary consists of words such as \textit{faculty}, \textit{kindergarten}, or \textit{phd}.
The \textit{gendered} vocabulary, on the other hand, consists of words that are most predictive of gender, which includes, apart from gender pronouns and words such as \textit{husband} and \textit{wife}, words like \textit{dance}, \textit{art}, or \textit{engineering}, which are not evidently gendered but highly correlated with the sensitive attribute.
Finally, we train two logistic regression models\footnote{We use logistic regression to ensure that explanations are faithful to the underlying model.} on a balanced set of professor and teacher bios, and we employ the \texttt{TextExplainer} from LIME~\citep{ribeiro2016should} to generate dynamic explanations with highlighting of predictive words.

\subsection{Measuring Reliance and Fairness}\label{subsec:measuring_reliance_fairness}

\paragraph{Selection of bios}
In order to be able to assess differences in reliance behavior across conditions, participants see a mix of cases where the AI is correct and where it is incorrect.
More specifically, we distinguish six types of scenarios that make up the 14 bios that participants see---they are summarized in \Cref{tab:scenarios}.
We distinguish these scenarios based on three dimensions: $(i)$ gender of the person associated with a bio; $(ii)$ true occupation of that person; $(iii)$ AI recommended occupation.
We show 3 cases each of correctly recommended women teachers (WTT) and men professors (MPP), as well as 3 cases of incorrectly recommended women professors (WPT) and men teachers (MTP).
Note that our focus is on scenarios where the AI recommendations are in line with gender stereotypes.
To preempt the misconception that the AI always recommends \emph{teacher} for women and \emph{professor} for men, we also include one case each of correctly recommended woman professor (WPP) and correctly recommended man teacher (MTT).
We include the WPP and MTT scenarios early on in our questionnaires.
Precisely, we randomize the order in which participants see the 14 bios, with the restriction that the WPP and MTT scenarios are shown among the first five---that way, we aim to avoid situations where participants see too many incorrect AI recommendations (in line with stereotypes) early on, which has been shown to negatively affect human reliance on AI systems~\citep{kim2020algorithms} and might, therefore, bias our results.
We do not consider scenarios where women teachers are classified as professors, or where men professors are classified as teachers, because our focus is on the errors that are more likely to occur in practice~\citep{de2019bias}.

\begin{table*}[t]
    \centering
    \caption{Our study includes 14 bios, consisting of three scenarios of types WTT, WPT, MTP, and MPP, respectively, and one scenario each of types WPP and MTT.}
    \label{tab:scenarios}
    \begin{tabular}{c c c c c c}
    \toprule
    \bf Gender of bio & \bf True occupation & \bf AI recommendation & \bf AI correct? & \bf Acronym & \bf \#Bios \\
    \midrule
    \textbf{W}oman & \textbf{T}eacher & \textbf{T}eacher & \cmark & WTT & 3\\
    \textbf{W}oman & \textbf{P}rofessor & \textbf{T}eacher & \xmark & WPT & 3\\
    \textbf{W}oman & \textbf{P}rofessor & \textbf{P}rofessor & \cmark & WPP & 1 \\
    \midrule
    \textbf{M}an & \textbf{T}eacher & \textbf{T}eacher & \cmark & MTT & 1 \\
    \textbf{M}an & \textbf{T}eacher & \textbf{P}rofessor & \xmark & MTP & 3\\
    \textbf{M}an & \textbf{P}rofessor & \textbf{P}rofessor & \cmark & MPP & 3\\
    \bottomrule
    \end{tabular}
\end{table*}

All bios shown to participants are taken from a random holdout set of BIOS that our two classifiers make predictions on. Specifically, we choose bios that are reasonably similar in length and where both classifiers yield the same predicted occupation as well as similar prediction probabilities.
We also require that these prediction probabilities for a bio must not be too high, which aims at eliminating bios that are ``too easy'' to classify.
The authors then manually screened the remaining contenders to settle on the final 14 bios.
The whole selection process is described in more detail in \Cref{sec:selection_of_scenarios}.

\paragraph{Measuring reliance behavior}
In our assessment of reliance behavior, we distinguish four cases, as depicted in \Cref{tab:adherence_overriding}.
We refer to cases where humans adhere to correct AI recommendations as \emph{correct adherence}, to cases where humans adhere to incorrect recommendations as \emph{detrimental adherence},\footnote{Other authors, such as \citet{vasconcelos2023explanations}, have also used the term \textit{over-reliance} for adhering to incorrect AI recommendations. However, as described by \citet{lockey2021review}, this term can also refer to a global behavior associated with over-trust that goes beyond the level of individual decisions. Therefore, in accordance with the more nuanced terminology of \citet{alessa2022role}, we intentionally use the more precise wording that better reflects the specific context of our discussion. With a similar argument, we use the term \textit{detrimental overriding} instead of \textit{under-reliance}.} to cases where humans override correct recommendations as \emph{detrimental overriding}, and to cases where humans override incorrect recommendations as \emph{corrective overriding}.
Note that the sum of shares of correct adherence and corrective overriding make up the final decision-making accuracy~\cite{schoeffer2023interdependence}.
This taxonomy is similar to the one proposed by \citet{liu2021understanding} for trust; however, we want to stress the difference between trust and reliance (see \Cref{sec:explanations_reliance}).
When comparing participants' reliance behavior across conditions, we compute and report the relative shares of any of these four types of reliance behavior on the 14 bios that participants see.

\begin{table}[h!]
    \centering
    \caption{We distinguish four types of reliance in AI-assisted decision-making: humans can adhere to or override correct AI recommendations, or they can adhere to or override incorrect AI recommendations.}
    \label{tab:adherence_overriding}
    \resizebox{\columnwidth}{!}{
        \begin{tabular}{l | c c}
        \toprule
        & \bf Human adherence to AI & \bf Human overriding of AI  \\
        \midrule
        \bf AI correct & Correct adherence & Detrimental overriding \\
        \bf AI incorrect & Detrimental adherence & Corrective overriding \\
        \bottomrule
        \end{tabular}
    }
\end{table}

\paragraph{Measuring distributive fairness}
To evaluate distributive fairness of decisions, we measure disparities in error rates across gender~\citep{barocas-hardt-narayanan,chouldechova2017fair}, which is closely linked to the ideas of \textit{equalized odds} and \textit{equal opportunity}~\citep{hardt2016equality}.
From a fairness perspective, the goal is to minimize such disparities, so as to equalize the burden of being misclassified and, as a result, being excluded from exposure to relevant opportunities between men and women.
We formalize these disparities as follows: let $FP_W$ be the share of incorrectly predicted women professors, i.e., women professors that are predicted to be teachers, and $FT_W$ the share of incorrectly predicted woman teachers. Similarly define $FP_M$ and $FT_M$ for men.
We can then quantify disparities in error rates as follows:
\begin{align*}
    \text{Error rate disparity (Teacher $\rightarrow$ Professor)} &= |FT_W-FT_M| \\
    \text{Error rate disparity (Professor $\rightarrow$ Teacher)} &= |FP_W-FP_M|,
\end{align*}
where we use the notation of ``Teacher $\rightarrow$ Professor'' to indicate teachers that are incorrectly predicted as professors, and vice versa for ``Professor $\rightarrow$ Teacher.'' If we assume that the occupation of \textit{professor} is associated with a higher societal status than \textit{teacher}, we may also refer to cases of ``Teacher $\rightarrow$ Professor'' as \textit{promotions}, and to ``Professor $\rightarrow$ Teacher'' as \textit{demotions}. This will be important in the discussion of our findings.

\begin{figure*}[t]
    \centering
    \begin{minipage}[t]{0.3\textwidth}
        \centering
        \includegraphics[width=\textwidth]{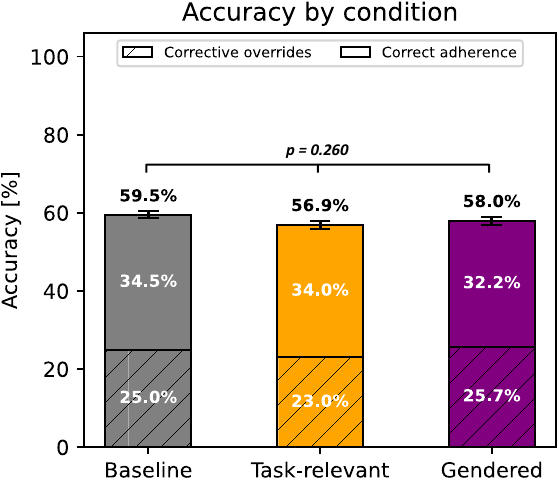}
        \caption{Accuracy is not higher when explanations are provided, compared to the baseline.}
        \label{fig:task_performance_by_condition}
        \Description{Accuracy is not higher when explanations are provided, compared to the baseline.}
    \end{minipage}\hfill
    \begin{minipage}[t]{0.3\textwidth}
        \centering
        \includegraphics[width=\textwidth]{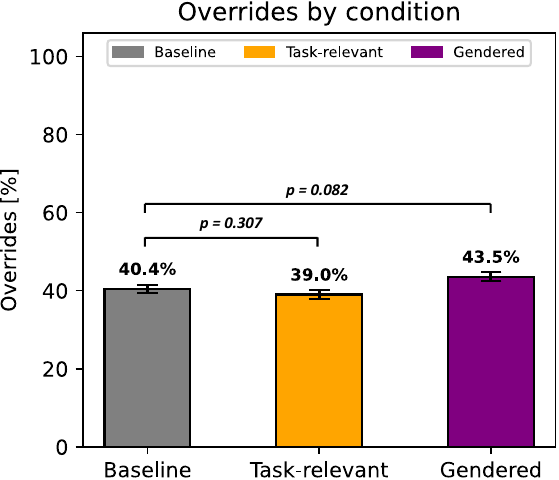}
        \caption{Overrides are highest in the \emph{gendered} condition.}
        \label{fig:overrides_by_condition}
        \Description{Overrides are highest in the gendered condition.}
    \end{minipage}\hfill
    \begin{minipage}[t]{0.3\textwidth}
        \centering
        \includegraphics[width=\textwidth]{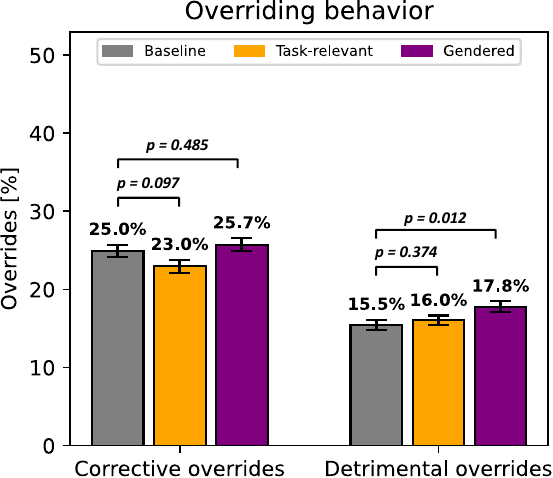}
        \caption{Explanations do not enable corrective vs. detrimental overrides.}
        \label{fig:overriding_behavior}
        \Description{Explanations do not enable corrective vs. detrimental overrides.}
    \end{minipage}
\end{figure*}

\paragraph{Measuring fairness perceptions}
To measure fairness perceptions, we provide a brief introduction and then ask participants' agreement with three statements, measured on 5-point Likert scales from 1 (``Fully disagree'') to 5 (``Fully agree'').
We operationalize this in our questionnaires similar to~\citet{colquitt2015measuring} as follows:
%
\begin{center}
\noindent
\fbox{%
    \parbox{0.9\columnwidth}{%
        \textit{The questions below refer to the procedures the AI uses to predict a person's occupation. Please rate your agreement with the following statements.
    \begin{enumerate}
        \item The AI's procedures are free of bias.
        \item The AI's procedures uphold ethical and moral standards.
        \item It is fair that the AI considers the highlighted words for predicting a person's occupation.
    \end{enumerate}}
    }%
}
\end{center}
%
Note that items (1) and (2) are taken from the \emph{procedural justice} construct of~\citet{colquitt2015measuring} and slightly rephrased to fit our case of AI-assisted decision-making.
These items have been frequently used in other human-AI studies, e.g., \citep{binns2018s,schoeffer2021study,schlicker2021expect,marcinkowski2020implications}.
Note that prior work has often measured fairness perceptions through single items only~\citep{starke2021fairness}.
\citet{colquitt2015measuring} propose up to eight measurement items for procedural justice in the organizational psychology context; however, several of these items are not applicable here.
Instead, we amend our questionnaires by a third item (3) that is more tailored to our experimental setup.
Since item (3) is more explicit and we want to avoid priming, we ask (3) last and without possibility to modify responses for (1) and (2) retroactively.
To obtain a single measure of fairness perceptions per participant, we eventually average ratings across the three items per participant; and we also confirm scale reliability in \Cref{sec:effects_proc_fairness}.

\subsection{Data Collection}\label{sec:data_collection}
Our study received clearance from an institutional ethics committee.
Participants were recruited via \texttt{Prolific}---a crowdworking platform for online research~\cite{palan2018prolific}.
We required participants to be at least 18 years of age, and to be fluent in English.
We also sampled approximately equal amounts of men and women; no other pre-screeners were applied.
After consenting to the terms of our study, participants were then randomly and in equal proportions assigned to one of our three conditions and asked to complete the respective questionnaire.
Overall, we recruited 600 lay people through \texttt{Prolific}.
At the time of taking the survey, 13.5\% of participants were 18--24 years old, 32.6\% were 25--34 years old, 21.3\% between 35--44, 13.8\% between 45--54, 11.3\% between 55--64, and 7.6\% were older than 65.
Regarding gender, 49.2\% identified as women, 48.0\% as men, and 1.8\% identified as non-binary / third gender, or preferred not to say.
8.0\% of participants are of Spanish, Hispanic, or Latinx ethnicity; and the majority (78.4\%) considered their race to be White or Caucasian, followed by Black or African American (7.0\%) and Asian (6.1\%).
For their participation, participants were paid on average \pounds10.58 (approximately \$12.70 at the time the study was conducted) per hour, excluding individual bonus payments of \pounds0.05 per correctly predicted occupation.
Participants took on average 10:12min (baseline), 12:51min (\emph{task-relevant}), and 12:27min (\emph{gendered}) to complete the survey.

\section{Analysis and results}\label{sec:results}
We first present results on the effects of explanations on accuracy as well as overriding behavior.
Then, we examine how reliance behavior translates to distributive fairness. Finally, we assess the role of fairness perceptions. For all statistical comparisons, we conduct nonparametric tests because we cannot confirm the prerequisites (normal distribution and equal variance) of their parametric counterparts.
Specifically, we conduct Kruskal-Wallis omnibus tests~\citep{kruskal1952use} whenever applicable, and two-tailed Mann-Whitney U tests~\citep{mann1947test} for pairwise comparisons. 
We report p-values for omnibus tests and pairwise comparison tests between the baseline and each condition with explanations (\textit{task-relevant} and \textit{gendered}), respectively, in the corresponding figures.

\subsection{Effects of Explanations on Accuracy and Overriding Behavior}\label{sec:effects_of_explanations}

\paragraph{Effects on accuracy}
First, we examine how accuracy may be different between the baseline and the conditions with explanations, \emph{task-relevant} and \emph{gendered}.
Mean accuracies\footnote{We use $M$ as a shorthand for \emph{mean}, and $SD$ for \emph{standard deviation}. We also use the subscripts  ${base}$, ${rel}$, and ${gen}$ to refer to the baseline, task-relevant, and gendered conditions, respectively.} per condition are $M_{base}=59.49\%$ ($SD_{base}=13.11$), $M_{rel}=56.94\%$ ($SD_{rel}=13.86$), and $M_{gen}=57.96\%$ ($SD_{gen}=14.30$), as shown in \Cref{fig:task_performance_by_condition}.\footnote{In figures we provide standard errors as error bars, where we compute the measure of interest (e.g., accuracy) for each individual participant in a given condition, then compute the standard deviation across all participants in that condition, and divide the result by the square root of the number of participants in that condition.}
The Kruskal-Wallis omnibus test further suggests that there are no significant differences between the three means ($p=0.260$).
Recall that participants were incentivized through bonus payments to accurately predict occupations.
This suggests that \textbf{explanations did not aid AI-assisted decision-making when measured in terms of accuracy.}

\begin{figure*}[t]
    \centering
    \begin{minipage}[t]{0.3\textwidth}
        \centering
        \includegraphics[width=\textwidth]{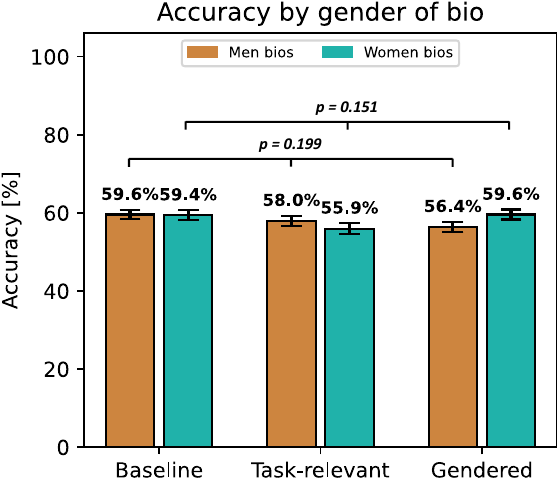}
        \caption{Explanations (middle and right) do not increase accuracy over the baseline, neither for men nor women bios.}
        \label{fig:acc_by_gender}
        \Description{Explanations do not increase accuracy over the baseline, neither for men nor women bios.}
    \end{minipage}\hfill
    \begin{minipage}[t]{0.3\textwidth}
        \centering
        \includegraphics[width=\textwidth]{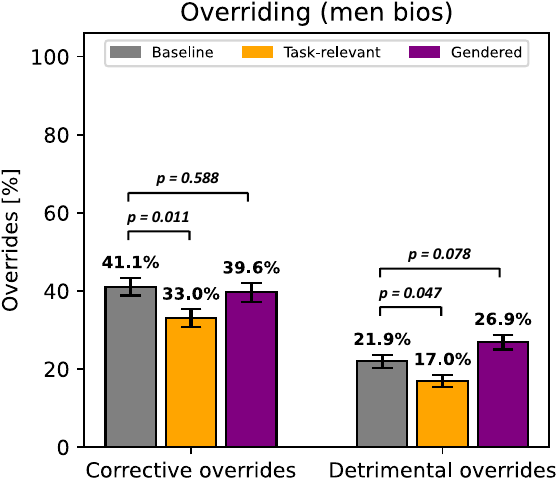}
        \caption{\emph{Task-relevant} explanations decrease both corrective and detrimental overrides for men bios, compared to the baseline; whereas \emph{gendered} explanations marginally increase detrimental overrides.}
        \label{fig:overriding_behavior_M}
        \Description{Task-relevant explanations decrease both corrective and detrimental overrides for men bios, compared to the baseline; whereas gendered explanations marginally increase detrimental overrides.}
    \end{minipage}\hfill
    \begin{minipage}[t]{0.3\textwidth}
        \centering
        \includegraphics[width=\textwidth]{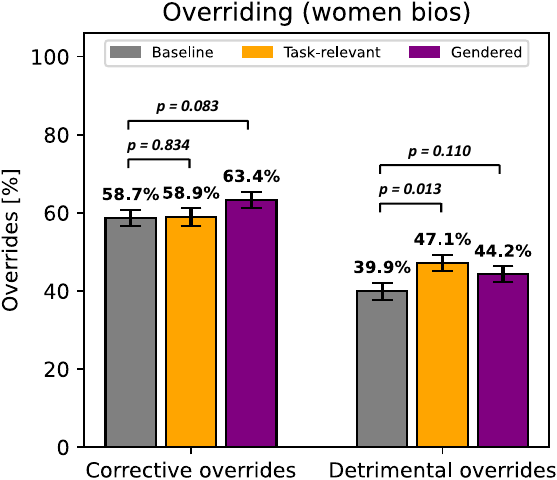}
        \caption{\emph{Gendered} explanations marginally increase corrective overrides over the baseline for women bios; whereas \emph{task-relevant} explanations increase detrimental overrides.}
        \label{fig:overriding_behavior_W}
        \Description{Gendered explanations marginally increase corrective overrides over the baseline for women bios; whereas task-relevant explanations increase detrimental overrides.}
    \end{minipage}
\end{figure*}

\paragraph{Effects on overriding behavior}
In \Cref{fig:overrides_by_condition,fig:overriding_behavior}, we see that participants in the \emph{gendered} condition overrode more AI recommendations than in the \emph{task-relevant} condition ($p=0.005$).
From \Cref{fig:overriding_behavior} we further conclude that \emph{both} corrective \emph{and} detrimental overrides are highest in the \emph{gendered} condition, with detrimental overrides being significantly higher than the baseline ($p=0.012$).
We interpret this increase in overrides further in \Cref{sec:interplay_explanations_reliance_fairness}.
In the \emph{task-relevant} condition, we see that overall overrides are lowest across conditions (\Cref{fig:overrides_by_condition}), with corrective overrides being marginally\footnote{We report \textit{marginal} significance for $0.05 < p \leq 0.10$ in line with prior work~\citep{olsson2019prevalence}, noting that such p-values have lower evidential value.} lower $(p=0.097)$ and detrimental overrides not significantly different $(p=0.374)$ compared to the baseline (\Cref{fig:overriding_behavior}).
Overall, we conclude that people's reliance behavior is affected by how the AI explains its recommendations; notably, people overrode AI recommendations more often when explanations highlight features that are evidently associated with gender.
Across conditions, we also infer from \Cref{fig:overriding_behavior} that participants generally performed more corrective than detrimental overrides, and that \textbf{the ability to perform corrective vs. detrimental overrides (i.e., the ratio of corrective to detrimental overrides) did not improve through the provision of explanations.}

\subsection{Interplay Between Explanations, Reliance, and Distributive Fairness}\label{sec:interplay_explanations_reliance_fairness}

\paragraph{Accuracy by gender}
Consistent with our findings at the aggregated level (see \Cref{fig:task_performance_by_condition}), we do not observe any accuracy changes through explanations over the baseline in \Cref{fig:acc_by_gender}, neither for men ($p=0.199$) nor women ($p=0.151$) bios.
This means that \textbf{both in the \emph{task-relevant} and the \emph{gendered} condition, explanations did not enable people to improve decision-making accuracy, neither for men nor women bios.}
From \Cref{fig:overriding_behavior_M,fig:overriding_behavior_W}, we see why this is the case: for \textit{men} bios, overrides overall went down in the \textit{task-relevant} condition, with corrective overrides decreasing even stronger than detrimental overrides; and overrides went \textit{up} in the \textit{gendered} condition, with only detrimental overrides increasing. For \textit{women} bios, detrimental overrides increased and corrective overrides did not change in the \textit{task-relevant} condition, whereas both corrective and detrimental overrides went up equally in the \textit{gendered} condition.

\paragraph{Types of overrides by gender and occupation}
When looking at effects of explanations on overriding behavior by gender in \Cref{fig:overriding_behavior_M,fig:overriding_behavior_W}, no intervention improved participants' ability to perform corrective vs. detrimental overrides of AI recommendations compared to the baseline---i.e., the ratio of corrective to detrimental overrides did not improve---neither for men nor women bios.
This is consistent with our findings at the aggregate level (see \Cref{fig:overriding_behavior}).
Notably, we see that detrimental overrides in the \emph{gendered} condition marginally increase for men bios (\Cref{fig:overriding_behavior_M}) over the baseline $(p=0.078)$, and in the \textit{task-relevant} condition they significantly increase for women bios (\Cref{fig:overriding_behavior_W}) compared to the baseline $(p=0.013)$.
At the same time, corrective overrides remain unchanged in either case.

From comparing \Cref{fig:overriding_behavior_M,fig:overriding_behavior_W}, it also appears that participants generally overrode more recommendations for women than men bios: both corrective and detrimental overrides are higher across all conditions in \Cref{fig:overriding_behavior_W} compared to \Cref{fig:overriding_behavior_M}.
However, because the AI system mostly predicts men as professors and women as teachers, \Cref{fig:overriding_behavior_M,fig:overriding_behavior_W} alone do not allow us to disentangle whether participants overrode women bios more often because of $(i)$ the bios' associated gender or $(ii)$ because of the fact that they were predicted to be teachers.
\Cref{fig:mtt,fig:wpp} in \Cref{sec:overrides_mtt_wpp} allow a more nuanced conclusion: here, we show that there are more overrides for men when they are correctly predicted by the AI model as teachers (MTT) than for women when they are correctly predicted by the AI model as professors (WPP).
Together, these results suggest that participants did not generally override women bios more often than men bios, but instead \textbf{people were overall more prone to do promoting\footnote{We assume here that the occupation of \emph{professor} is associated with a higher societal status than that of \emph{teacher}. Hence, \emph{promoting} refers to predicting someone to be a professor, whereas \emph{demoting} means to predict someone to be a teacher.} overrides}; which means that participants overrode AI recommendations more often when someone was suggested to be a teacher vs. a professor.

\begin{figure*}[t]
    \centering
    \begin{minipage}[t]{0.3\textwidth}
        \centering
        \includegraphics[width=\textwidth]{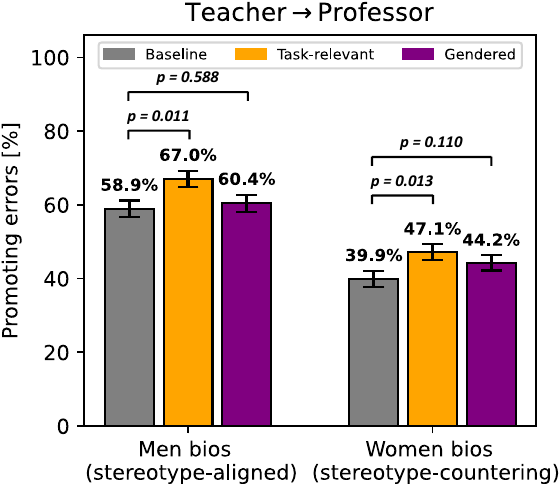}
        \caption{Bios incorrectly classified by humans as professor: Promoting errors increase for both men and women bios in the \emph{task-relevant} condition, compared to the baseline.}
        \label{fig:fp}
        \Description{Bios incorrectly classified by humans as professor: Promoting errors increase for both men and women bios in the task-relevant condition, compared to the baseline.}
    \end{minipage}\hfill
    \begin{minipage}[t]{0.3\textwidth}
        \centering
        \includegraphics[width=\textwidth]{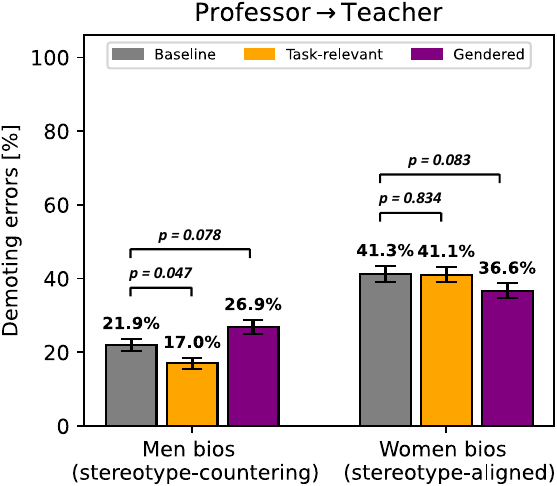}
        \caption{Bios incorrectly classified by humans as teacher: In the \emph{gendered} condition, demoting errors marginally increase for men bios (left) and decrease for women bios (right), compared to the baseline; and they only decrease for men bios in the \emph{task-relevant} condition.}
        \label{fig:fn}
        \Description{Bios incorrectly classified by humans as teacher: In the gendered condition, demoting errors marginally increase for men bios and decrease for women bios, compared to the baseline; and they only decrease for men bios in the task-relevant condition.}
    \end{minipage}\hfill
    \begin{minipage}[t]{0.3\textwidth}
        \centering
        \includegraphics[width=\textwidth]{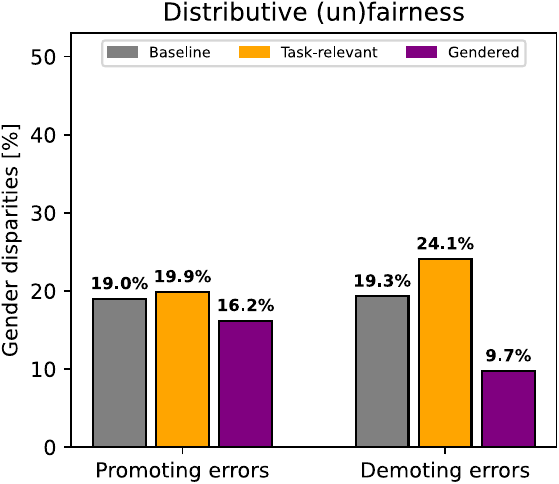}
        \caption{Absolute error differences between men and women: \emph{Gendered} explanations decrease both disparities in promoting (teacher $\rightarrow$ professor) and demoting (professor $\rightarrow$ teacher) errors between genders, compared to the baseline; \emph{task-relevant} explanations increase disparities.}
        \label{fig:gaps}
        \Description{Absolute error differences between men and women: Gendered explanations decrease both disparities in promoting and demoting errors between genders, compared to the baseline; task-relevant explanations increase disparities.}
    \end{minipage}
\end{figure*}

Importantly, people's likelihood to override conditioned on gender and predicted occupation did vary across conditions.
By virtue of our study design, we are able to observe stereotype-countering\footnote{Recall that societal stereotypes typically associate men with being professors and women with being teachers~\citep{miller2000women}.} corrective overrides, and both stereotype-aligned and stereotype-countering detrimental overrides.
As explained in \Cref{sec:study_setup}, the motivation for this design is our focus on studying whether explanations allow humans to correct for stereotype-aligned incorrect AI predictions, which would be the most frequent errors of an occupation prediction model that exhibits gender bias~\citep{de2019bias}.
We see that in the \emph{task-relevant} condition, people performed fewer corrective overrides for men ($p=0.011$) and the same amount for women ($p=0.834$) in comparison to the baseline, as shown in \Cref{fig:overriding_behavior_M,fig:overriding_behavior_W}.
Meanwhile, in the \emph{gendered} condition participants performed marginally more corrective overrides for women ($p=0.083$) and the same amount of such overrides for men ($p=0.588$).
This means that \textbf{participants in the \emph{gendered} condition were more likely to perform stereotype-countering corrective overrides than in the baseline, while participants in the \emph{task-relevant} condition were less likely to do so.}

As for detrimental overrides, we see that they marginally increase in the \emph{gendered} condition for men bios ($p=0.078$), compared to the baseline (\Cref{fig:overriding_behavior_M}).
We also see that they are higher for women bios in the \textit{gendered} condition (\Cref{fig:overriding_behavior_W}), even though not statistically significant ($p=0.110$).
Considering that we do not observe differences in stereotype-aligned detrimental overrides between conditions (\Cref{fig:mtt,fig:wpp} in \Cref{sec:overrides_mtt_wpp}), we infer that people in the \emph{gendered} condition performed more stereotype-countering detrimental overrides, by predicting more men to be teachers and women to be professors.
It is noteworthy that when contrasting corrective and detrimental overrides, we observe that \textbf{no condition improved participants' ability to make stereotype-countering \textit{corrective} overrides vs. stereotype-countering \textit{detrimental} overrides}.
In the \textit{gendered} condition, this means that participants became more likely to override an AI recommendation when it predicted that a woman is a teacher, irrespective of her true occupation.
\textbf{Overall, we observe reliance behavior in the \textit{gendered} condition that counters societal stereotypes, whereas in the \textit{task-relevant} condition people tend to rely on AI recommendations in a way that reinforces stereotypes.}
We elaborate on the implications of this for distributive fairness below.

\paragraph{Implications for distributive fairness}
We now examine how the observed reliance behavior relates to distributive fairness with respect to disparities in errors between men and women.
First, we note that in the baseline condition, people tend to make more errors that promote men vs. women (58.9\% vs. 39.9\% in \Cref{fig:fp}), and erroneously demote women more than men (41.3\% vs. 21.9\% in \Cref{fig:fn}).
Note that in the case of men, promoting behavior is stereotype-aligned, whereas in the case of women such behavior is stereotype-countering; and vice versa for demoting behavior.
The resulting absolute error rate disparities between men and women for the baseline are, hence, 19.0\% (promotions) and 19.3\% (demotions), as depicted in \Cref{fig:gaps}.
From the previous paragraph we know that people in the \emph{task-relevant} condition showed a tendency of reinforcing stereotypes, meaning that promotions of men increased more than those of women, which increased disparities in promotions even further over the baseline (\Cref{fig:gaps}, left).
Similarly, demotions of men decreased much more than demotions of women, leading to increased disparities in demotions over the baseline (\Cref{fig:gaps}, right).
In conclusion, we note that \textbf{people's stereotype-aligned reliance behavior in the \emph{task-relevant} condition exacerbated existing disparities in the baseline condition and, hence, hindered distributive fairness.}

\begin{figure*}
    \centering
    \begin{minipage}[t]{0.475\textwidth}
        \centering
        \includegraphics[width=\textwidth]{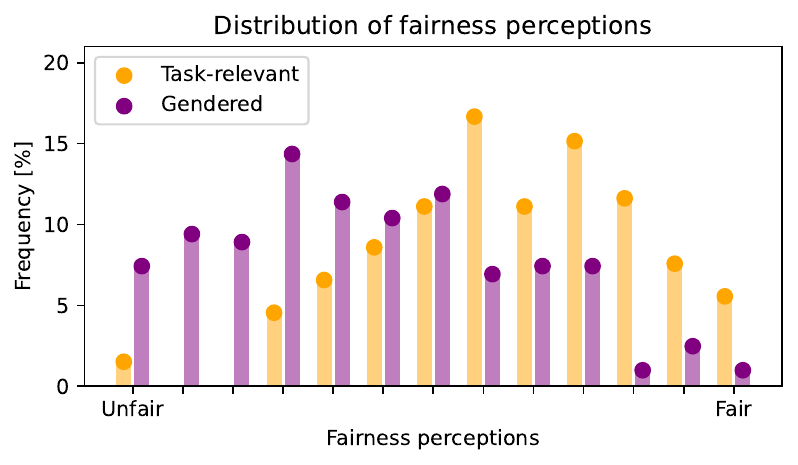}
        \caption{Fairness perceptions are higher in the \emph{task-relevant} condition compared to the \emph{gendered} condition. Fairness perceptions are averages of three items measured on 5-point Likert scales, resulting in values between 1 (``unfair'') and 5 (``fair'') with 0.33 increments.}
        \label{fig:perceptions}
        \Description{Fairness perceptions are higher in the task-relevant condition compared to the gendered condition.}
    \end{minipage}\hfill
    \begin{minipage}[t]{0.475\textwidth}
        \centering
        \includegraphics[width=\textwidth]{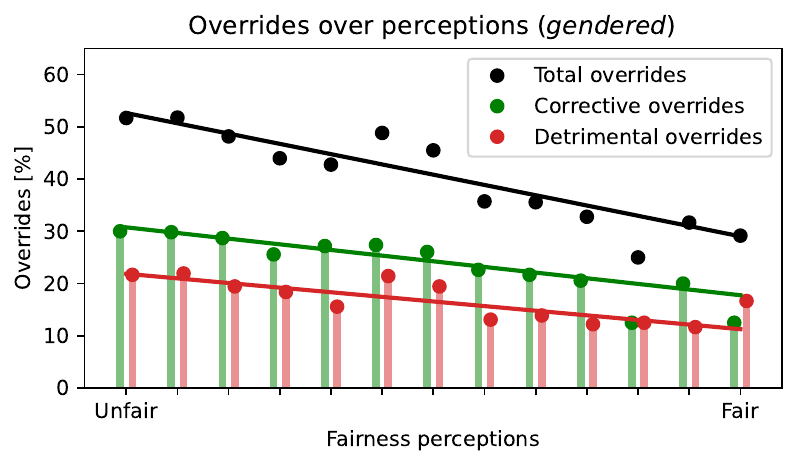}
        \caption{Significant negative relationship between fairness perceptions and overrides, both corrective and detrimental, as well as overall. Ratio of corrective to detrimental overrides is independent of fairness perceptions.}
        \label{fig:overrides_over_perceptions_gendered}
        \Description{Significant negative relationship between fairness perceptions and overrides, both corrective and detrimental, as well as overall. Ratio of corrective to detrimental overrides is independent of fairness perceptions.}
    \end{minipage}
\end{figure*}

In the \emph{gendered} condition, on the other hand, people countered stereotypes, meaning that promotions of women increased more than for men, reducing existing disparities (\Cref{fig:gaps}, left).
The most drastic reduction in disparities happens for demotions (\Cref{fig:gaps}, right), since demotions \emph{increased} for men and \emph{decreased} for women (\Cref{fig:fn}).
This results in a reduction of disparities in demotions from 19.3\% (baseline) to 9.7\% (\emph{gendered} condition).
Hence, \textbf{people's stereotype-countering reliance behavior in the \emph{gendered} condition mitigated existing disparities and, hence, fostered distributive fairness.}
It is important to stress that while disparities in error types decreased in the \emph{gendered} condition compared to the baseline, this was due to a shift in the types of errors, as opposed to an increased ability to override mistaken AI recommendations.

\subsection{The Role of Fairness Perceptions}\label{sec:effects_proc_fairness}

\paragraph{Effects of explanations on fairness perceptions}
Recall that we measure three items regarding fairness perceptions on 5-point Likert scales, ranging from 1 (unfair) to 5 (fair), as outlined in \Cref{subsec:measuring_reliance_fairness}.
We confirm good scale reliability at a Cronbach's alpha~\citep{taber2018use} value of 0.77.
We then take the average of the three item ratings for each participant to obtain a single measure of fairness perceptions.
From the distribution in \Cref{fig:perceptions}, we see that participants in the \emph{task-relevant} and \emph{gendered} conditions have significantly different perceptions of fairness towards the AI model.
Concretely, we observe $M_{rel}=3.53$ ($SD_{rel}=0.85$) in the \emph{task-relevant} condition, and $M_{gen}=2.54$ ($SD_{gen}=0.98$) in the \emph{gendered} condition.
This means that people who are shown a highlighting of task-relevant words perceived the underlying AI as fairer than people who were shown gendered words as being important for given AI recommendations.
Overall, we confirm prior works' findings and conclude that \textbf{the AI system was perceived as significantly less fair when explanations point at the use of sensitive features compared to cases where explanations point at task-relevant features.}
Since it is known that explanations can be engineered to obfuscate the use of of sensitive features by an AI system~\citep{dimanov2020you,lakkaraju2020fool,pruthi2019learning,slack2020fooling}, these findings imply that explanations can be exploited to mislead people's perceptions in the spirit of what \citet{aivodji2019fairwashing} have been referring to as ``fairwashing.''

\paragraph{Relationship of fairness perceptions with overriding behavior}
When we look at people's overriding behavior as a function of their fairness perceptions, we observe a strong negative relationship ($p=1.10\times 10^{-11}$) between fairness perceptions and overriding of AI recommendations; i.e., participants overrode the AI more often when their fairness perceptions were lower.
Concretely, we see that people overrode on average 52\% of AI recommendations when their fairness perceptions were lowest, and only 31\% when their fairness perceptions were highest.
This negative relationship is consistent in both the \emph{task-relevant} and the \emph{gendered} condition, and it also persists when we disentangle corrective and detrimental overrides at the aggregate level.
\Cref{fig:overrides_over_perceptions_gendered} shows the relationship of overrides---both corrective, detrimental, and total---as a function of fairness perceptions for the \emph{gendered} condition.
Dots represent mean values of overrides for a given level of perceptions, and lines are OLS regressions fitted on the original data.
All slopes in \Cref{fig:overrides_over_perceptions_gendered} are significantly negative (total: $p=1.97\times 10^{-7}$; corrective: $p=9.18\times 10^{-5}$; detrimental: $p=1.53\times 10^{-4}$).
We observe that as participants overrode more AI recommendations in the \emph{gendered} condition, the rates at which corrective and detrimental overrides increase are approximately equal---in other words, the ratio of corrective to detrimental overrides is constant across perceptions.
Overall, we conclude that \textbf{people's fairness perceptions are associated with their reliance behavior in a way that low perceptions relate to more overrides than high perceptions.
However, both corrective \emph{and} detrimental overrides increased as fairness perceptions decreased.}
This implies that perceptions are not an indicator of people's ability to perform corrective vs. detrimental overrides, but tend to only be associated with the quantity of overrides.
Concerningly, this also means that explanations may not only be exploited to mislead people's fairness perceptions but also their reliance behavior---in a way that people override more AI recommendations when explanations indicate the use of sensitive information, compared to the case where only task-relevant features are highlighted.

\section{Discussion}
In this section, we discuss our findings more broadly.
We begin by summarizing our main findings, then acknowledge limitations of our study, and finally provide implications and recommendations for the design of socio-technical systems for decision support.
In this section, we also discuss several directions for future work.

\subsection{Summary of Findings}
In this work, we conducted a first holistic analysis of the effects of feature-based explanations on distributive fairness in AI-assisted decision-making.
We also studied the mediating roles of reliance behavior and fairness perceptions, which have been the focus of prior work.
Our findings suggest that feature-based explanations can have different effects on people's perceptions, their reliance behavior, and distributive fairness---depending on whether they highlight the use of task-relevant words vs. words that are proxies for sensitive attributes.
Specifically, we observe that for the task of occupation classification based on bios, a highlighting of gendered words led to lower fairness perceptions, which are associated with more overrides of AI recommendations.

These findings relate to \textit{algorithm aversion theory}~\citep{dietvorst2015algorithm}, which postulates that humans tend to reject advice from algorithms (e.g., AI systems) and instead favor human judgment after they see the algorithm err.
\citet{jussupow2020we} frame algorithm aversion more generally as ``negative behaviours and attitudes towards the algorithm,'' and discuss several reasons for it, including algorithm agency, performance/capabilities, and human involvement.
Our findings might be seen as an extension to this theory as they suggest perceived unfairness as another driver of human aversion towards AI systems.
These findings can also be related to recent work by \citet{vasconcelos2023explanations}, who argue that the effects of explanations on reliance behavior depend on the degree to which they reduce the cognitive effort of verifying AI recommendations. 
While we do not observe that explanations \textit{actually} enabled humans to verify the correctness of AI recommendations, a speculative reason why explanations in the \textit{gendered} condition triggered more overrides is that they enabled participants to easily see that the AI system was considering sensitive information, which then resulted in low fairness perceptions and an increase in overrides.
On the other hand, when task-relevant words are highlighted, this led to higher fairness perceptions, which translate to fewer overrides.
In no case, however, do we observe that explanations improved people's ability to perform corrective vs. detrimental overrides, compared to a scenario with no explanations.

Finally, we show that feature-based explanations can improve \textit{or} hinder distributive fairness by fostering shifts in errors that counter or reinforce stereotypes: in the \textit{gendered} condition, participants displayed stereotype-countering reliance behavior, while in the \textit{task-relevant} condition, they displayed stereotype-aligned behavior. 
In both these cases, the respective reliance behavior affected both corrective and detrimental overrides.
This means that the conditions affected the likelihood to perform an override conditioned on the predicted occupation and a bio's associated gender, but with no relationship to the true occupation.
For instance, the \textit{gendered} condition fostered \textit{more} overrides of AI recommendations when a woman was predicted to be a teacher, irrespective of whether this prediction was correct; meanwhile, in the \textit{task-relevant} condition participants were \textit{less} likely to override AI recommendations where a man was predicted to be a professor, irrespective of his true occupation.

\subsection{Limitations}
Our study setup assigned participants to either the \emph{gendered} or the \emph{task-relevant} condition; i.e., participants saw either only explanations with highlighting of gendered words or task-relevant words.
We made this choice because we wanted to measure perceptions of fairness, but eliciting perceptions at an instance level could lead people to anchor their decisions to their expressed perceptions (or vice versa), which would compromise external validity.
Assigning people to different conditions enabled us to measure perceptions at the aggregate level.
In practice, an AI model might sometimes highlight only sensitive features, sometimes only task-relevant features, and at other times a mix of both.
Future work that studies how instance-level perceptions relate to aggregate-level perceptions, and how these interdependencies shape reliance behavior could complement our findings.
While our study design does not explicitly account for this, even if perceptions vary at the instance level, our findings suggest that reliance would depend on the inclusion of sensitive features, which research has shown to be an unreliable signal for assessing algorithmic fairness~\citep{kleinberg2018algorithmic,apfelbaum2010blind,dwork2012fairness,nyarko2021breaking,pedreshi2008discrimination,lakkaraju2020fool,pruthi2019learning}.
In particular, previous research has shown that ``fairness through unawareness,'' i.e., the exclusion of information that is evidently indicative of a person's demographics, is neither necessary nor sufficient for an algorithm to be procedurally fair~\citep{lakkaraju2020fool,pruthi2019learning,nyarko2021breaking} or to not display bias in terms of distributive fairness~\citep{dwork2012fairness,kleinberg2018algorithmic,pedreshi2008discrimination,apfelbaum2010blind}.
Our paper complements these works by showing that feature-based explanations may foster stereotype-aligned reliance behavior, therefore \emph{hindering} distributive fairness in AI-assisted decisions.

Importantly, our study does not claim that the observed effects should be taken for granted in any AI-assisted decision-making setup; rather, they should be carefully evaluated in future empirical studies.
With this work, we aim to provide an important example that shows how unreliable feature-based explanations are when it comes to effects on humans' reliance behavior and distributive fairness.
Our hope is that this work will inform improved assessment and design of explanability techniques, leading to a nuanced understanding of when and how certain types of explanations can enable humans to improve fairness properties of a system. 

\subsection{Implications}
Our work has several important implications for the design of socio-technical systems for decision support.
In the following, we group them by theme.

\paragraph{Towards meaningfully evaluating explanations}



A main argument of our work is that claims around explanations fostering distributive fairness must directly measure the impact of explanations on fairness metrics of AI-assisted decisions, which depend on humans' reliance behavior.
To this end, our study constitutes a blueprint that should be used to evaluate other types of explanations and tasks.
Crucially, our research shows that the mechanism through which reliance behavior affects metrics of fairness matters.
In particular, we show that distributive fairness may improve even in the absence of an enhanced ability to perform corrective overrides.
In other words, appropriate reliance is not a necessary condition for improving distributive fairness in AI-assisted decision-making, but the presence of explanations may drive a change in fairness metrics by fostering over- or under-reliance for certain types of cases.
Simultaneously, our work shows that an improvement in distributive fairness metrics does not necessarily mean that humans are overriding \emph{incorrect} recommendations.
For instance, we have seen in our study that overriding \textit{stereotype-aligned} AI recommendations (e.g., when a woman is predicted to be a teacher) 
may decrease gender disparities in error rates, even if correct recommendations are being overridden.
This finding may be particularly important from a design and a policy perspective, since a common motivation when providing humans with discretionary power to override decisions is an expectation that they will be able to correct for an AI system's mistakes~\cite{de2020case,GDPR}. 

These findings also have implications for the interpretation of studies focused on perceptions of fairness.
Our work shows that fairness perceptions have no bearing on people's ability to correctively override AI recommendations.
Instead, our study results suggest that low fairness perceptions are associated with more overrides of AI recommendations, irrespective of their correctness.
This may still lead to improvements in distributive fairness but does not indicate that humans differentiate between correct and incorrect AI recommendations.
This is important as perceptions are often used as proxies for trust and reliance~\citep{starke2021fairness}.


\paragraph{Towards grounding explanations in concrete and realistic objectives}
Previous work has emphasized that interpretability is not a monolithic concept, and the design of explanations should always be grounded on a concrete objective that it helps advance~\citep{lipton2018mythos}.
However, as also noted in previous research~\citep{deck2023critical,langer2021we}, it appears that we have been overloading explanations with a plethora of vague and broad-brush expectations---especially in regards to fairness---which may promote a false sense of optimism with respect to the capabilities of existing techniques, and hinders the design of truly effective novel interventions.
Prior work has claimed, e.g., that ``explainability can be considered as the capacity to reach and guarantee fairness in ML models''~\citep{arrieta2020explainable}. However, such claims are misleading as they ignore the fact that both explainability and fairness are complex, multi-dimensional concepts that warrant a more nuanced perspective~\citep{deck2023critical}.
%
Our work emphasizes the importance of designing explanations with the explicit purpose of enabling human decision-makers to rely on AI recommendations in a way that enhances distributive fairness metrics, and it casts doubt over the reliability of popular feature-based explainability approaches to advance this goal---despite them being widely employed in both academic and practical settings~\citep{bhatt2020explainable,gilpin2018explaining}.

\paragraph{Towards providing relevant cues through explanations}

Related to the previous theme, we argue that explanations must be designed to provide relevant cues to human stakeholders for them to be able to use their discretionary power effectively towards advancing any specified objectives.
With respect to distributive fairness, our work suggests that feature-based explanations are not providing these relevant cues.
Informing humans about whether or not an AI system considers sensitive information (e.g., gender) would only be relevant if it were desirable to override AI recommendations based thereon.
However, previous research has shown that the disuse of sensitive information (``fairness through unawareness''~\citep{kusner2017counterfactual}) is neither a necessary nor a sufficient condition for distributive fairness~\citep{corbett2018measure,dwork2012fairness,kleinberg2018algorithmic,nyarko2021breaking,pedreshi2008discrimination}.
Moreover, it is known that the use of sensitive information by an AI system can be concealed:
\citet{lakkaraju2020fool}, e.g., construct misleading explanations by leveraging correlations between sensitive and seemingly legitimate features.
This means that feature-based explanations are not a reliable mechanism to assess either procedural or distributive fairness.
Instead, we propose that explanations must transcend a human-in-the-loop operationalization of the flawed idea of ``fairness through unawareness'' and instead enable the human decision-maker to ground their reliance behavior on information that is both \textit{relevant} and \textit{reliable} for improving distributive fairness metrics.
A possible way forward might be to directly communicate relevant individual or group fairness properties of the AI system to the decision-maker, which has been recently studied by \citet{ashktorab2023fairness}.

%

\paragraph{Towards widening the scope of explanations}

We also suggest moving towards a broader understanding of what algorithmic transparency might entail. We should aim to understand how to build a supportive ecosystem around AI systems, one that enables relevant human stakeholders to achieve their respective goals. One element of such an ecosystem could be an interface that allows individuals to query different pieces of information, based on their background and situational needs.
To this point, novel findings from ethnographic work studying the use of AI have the potential to inform alternative designs of explanations.
For instance, ~\citet{lebovitz2022engage} study the adoption of AI in three healthcare domains and emphasize the importance of \emph{interrogation practices}, which are practices used by humans to relate their own knowledge to AI's predictions.
They note that if AI systems are to add value, they will sometimes make recommendations that conflict with experts’ knowledge. Therefore, what is needed are processes and tools that assist them in reconciling these differing perspectives.
Moreover, it is not always clear that what is needed are explanations pertaining to the AI system’s inner workings, as opposed to explanations of the broader socio-technical system. For instance, interventions that assist humans in reasoning about the information that is and is not available to the AI system may help them reconcile disagreements and better integrate multiple information sources~\citep{hemmer2022effect,holstein2022toward}. In clinical decision-making, \citet{ehsan2023charting} discovered that explanations could foster social interactions and reveal how different physicians responded to specific AI recommendations in the past.
Auxiliary interventions such as cognitive forcing functions have also been demonstrated to encourage more effective reliance behavior~\citep{buccinca2021trust}.


\paragraph{Additional future directions}

To design effective interventions for decision support, it is important to understand the psychological mechanisms at play when humans adhere to or override AI recommendations.
One promising direction for follow-up work will be to study \textit{why} the highlighting of gendered features results in AI aversion.
On the other hand, we have also seen cases where humans perceive the use of gendered words for predicting occupations as \textit{fair} (see \Cref{fig:perceptions} in \Cref{sec:results}), and it will be interesting to analyze when and why this is this case.
%

Prior work has argued that people's socio-cultural identity may relate to how they integrate AI recommendations into their decisions~\citep{fazelpour2022diversity}.
Empirically, \citet{mallari2020look} found that self-reported gender affected decision-making behavior and also interacted with the demographics of decision-subjects in the realm of recidivism prediction.
Moreover, \citet{peng2019you} provide evidence that crowdworkers' gender affected AI-assisted decisions, including biases.
Studying how demographics influence the use of explanations with respect to distributive fairness is, however, an open and important research question that merits in-depth follow-up work.

\section{Conclusion}
Explanations have been framed as an important mechanism for better and fairer human-AI decision-making.
In the context of fairness, however, this has not been appropriately studied, as prior works have mostly evaluated explanations based on their effects on people’s perceptions.
To fill this gap, we conducted a first comprehensive study of the effects of popular feature-based explanations on distributive fairness in AI-assisted occupation prediction.
We find that the type of features that an explanation highlights matters: when explanations highlight only task-relevant words, people tend to \textit{reinforce} stereotypical AI recommendations, ultimately \textit{increasing} error rate disparities between women and men. On the other hand, when explanations highlight gendered words, people tend to override more AI recommendations to \textit{counter} stereotypical AI recommendations, which \textit{decreases} error rate disparities.
Importantly, these effects on distributive fairness do not involve an enhanced human ability to override incorrect AI recommendations (i.e., ``appropriate reliance'') but solely emerge from a shifting in error types. For instance, if an AI system predicts that a woman is a teacher and the explanation highlights the use of gendered words, human decision-makers are more likely to override the recommendation regardless of whether the woman is indeed a teacher.

Overall, our findings raise doubts about the reliability of feature-based explanations as a mechanism to improve distributive fairness in AI-assisted decision-making. Our work has important implications for the design of socio-technical systems for decision support, pertaining to $(i)$ meaningfully evaluating explanations, $(ii)$ grounding explanations in concrete objectives, $(iii)$ providing relevant cues through explanations, and $(iv)$ widening the scope of explanations.
We hope that this work will inform improved assessment and design of novel explainability techniques, leading to a nuanced understanding of when and how certain types of explanations can enable humans to improve fairness properties of a system.

\begin{acks}
This research was supported in part by NIH grant R01NS124642, by a Google Award for Inclusion Research, and by {\em Good Systems\,}\footnote{\scriptsize\url{http://goodsystems.utexas.edu/}}, a UT Austin Grand Challenge to develop responsible AI technologies.
\end{acks}

\bibliographystyle{ACM-Reference-Format}
\bibliography{references}

\appendix

\section{Construction of task-relevant and gendered classifiers}\label{app:construction_tr_g}

Here, we explain in more detail how we constructed the AI models that we use for generating recommendations and explanations in the \emph{task-relevant} and \emph{gendered} conditions.

Let $\mathcal{W} \coloneq \{w_1,\dots,w_n\}$ be the set of $n$ words that occur most often across the set of all bios.
We chose $n=5000$, i.e., $\mathcal{W}$ contains the top-5000 most occurring words, after removal of (manually defined) stop words.
We inferred $\mathcal{W}$ from applying a \texttt{CountVectorizer}~\citep{scikit-learn}.
In trial runs, we found that increasing $n$ beyond 5000 does not significantly change the classifiers' predictions.
We then constructed two logistic regression classifiers, $\mathbf{AI}_{rel}$ and $\mathbf{AI}_{gen}$, with access to mutually disjoint vocabularies: \emph{task-relevant words} ($\mathcal{W}_{rel} \subset \mathcal{W}$) and \emph{gendered words} ($\mathcal{W}_{gen} \subset \mathcal{W}$).

\paragraph{Task-relevant vocabulary}
We performed the following steps to construct the task-relevant vocabulary $\mathcal{W}_{rel}$:

\begin{enumerate}
    \item For all $i\in\{1,\dots,n\}$, compute the average occurrence of word $w_i\in \mathcal{W}$ in bios of men and women professors and teachers. We call the results $\widehat{w_i^{P,m}}$, $\widehat{w_i^{P,w}}$, $\widehat{w_i^{T,m}}$, and $\widehat{w_i^{T,w}}$, where we use $P,T$ and $m,w$ as a shorthand for the respective occupations and genders. We also compute $\widehat{w_i^{\bullet}}$ as the average occurrence of $w_i$ for any other occupation~$\bullet$ that is \emph{not} professor or teacher.
    \item For given gender $g\in\{m,w\}$, check whether $\widehat{w_i^{P,g}} > \widehat{w_i^{\bullet}}$ \emph{or} $\widehat{w_i^{T,g}} > \widehat{w_i^{\bullet}}$ for all other occupations~$\bullet$, i.e., whether the average occurrence of word $w_i$ in professor or teacher bios of gender $g$ is greater than the average in \emph{any} other occupation. If this condition is met, add $w_i$ to $\mathcal{W}_{rel}^g$, the set of task-relevant words for gender $g$.
    \item Compute $\mathcal{W}_{rel}^{m} \cap \mathcal{W}_{rel}^{w} = \mathcal{W}_{rel}$ as the set of words that are task-relevant for \emph{both} genders.
\end{enumerate}
After completing steps (1)--(3), we obtain the task-relevant vocabulary $\mathcal{W}_{rel}$ of 543 words, including \emph{faculty}, \emph{kindergarten}, or \emph{phd}, among others.

\paragraph{Gendered vocabulary}
Denote $|\mathcal{B}^{o,g}|$ the amount of bios of occupation $o\in\{P,T\}$ and gender $g\in\{m,w\}$.
We perform the following steps to construct the gendered vocabulary $\mathcal{W}_{gen}$:

\begin{enumerate}
    \item Sample equal amounts of bios for men and women professors and teachers. Since $\min\{|\mathcal{B}^{o,g}|\}=|\mathcal{B}^{T,m}|=6440$, randomly sample 6440 bios for each combination of occupation and gender.
    \item Extract features from bios by applying a \texttt{CountVectorizer} with \texttt{TF-IDF} weighting~\citep{scikit-learn}.
    \item Train a logistic regression to predict \emph{gender} from the extracted features.
    \item Compute the importance of each (weighted) feature based on the absolute magnitude of their corresponding regression coefficient, and sort the resulting list of words by importance.
    \item Include the top-5\% most important words in $\mathcal{W}_{gen}$ as the set of words that are highly predictive of gender. We choose the threshold of 5\% so as to exclude words that are spuriously correlated with gender (e.g., \emph{towards}).
\end{enumerate}
After completing steps (1)--(5), we obtain the gendered vocabulary $\mathcal{W}_{gen}$ of 214 words, which include---apart from gender pronouns and words such as \emph{husband} and \emph{wife}---words like \emph{dance}, \emph{art}, or \emph{engineering}, which are not evidently gendered.

\paragraph{Deploying the classifiers}
Having established our vocabularies $\mathcal{W}_{rel}$ and $\mathcal{W}_{gen}$, we proceed by training two logistic regression models on a balanced set of bios containing 50\% professors and 50\% teachers.
Denote $|\mathcal{B}^P|$ and $|\mathcal{B}^T|$ the amounts of bios of occupations $P$ and $T$.
Since $|\mathcal{B}^T|=16,221<|\mathcal{B}^P|$, we randomly sample 16,221 bios of professors, while preserving the gender distribution from the original data.
This yields a dataset of 32,442 bios, 50\% of which we use as a holdout set.
We separate a relatively large holdout set because we will eventually use a specific subset of these bios in our questionnaires (see \Cref{sec:selection_of_scenarios}).
The resulting classifiers achieve $F_1$ scores of 0.87 ($\mathbf{AI}_{rel}$) and 0.77 ($\mathbf{AI}_{gen}$).
For generating dynamic explanations with highlighting of predictive words, we employ the \texttt{TextExplainer} from LIME~\citep{ribeiro2016should}.

\begin{figure*}[ht]
    \centering
    \begin{minipage}[t]{0.475\textwidth}
        \centering
        \includegraphics[width=0.8\textwidth]{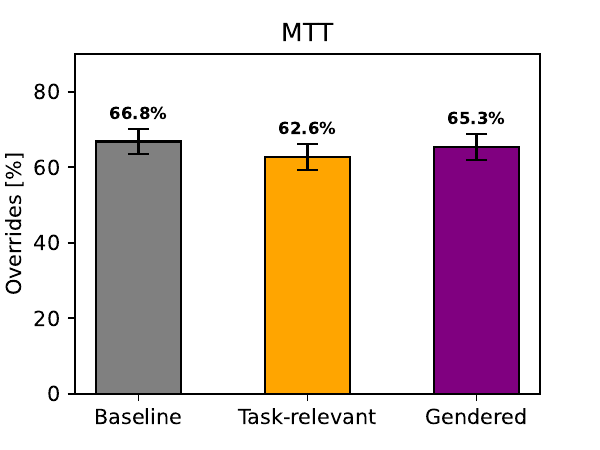}
        \caption{Overrides of AI recommendations that correctly predict a man teacher to be a teacher (MTT).}
        \label{fig:mtt}
        \Description{Overrides of AI recommendations that correctly predict a man teacher to be a teacher (MTT).}
    \end{minipage}\hfill
    \begin{minipage}[t]{0.475\textwidth}
        \centering
        \includegraphics[width=0.8\textwidth]{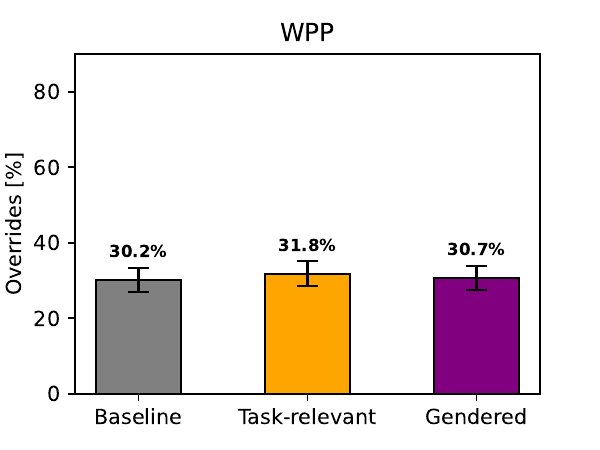}
        \caption{Overrides of AI recommendations that correctly predict a woman professor to be a professor (WPP).}
        \label{fig:wpp}
        \Description{Overrides of AI recommendations that correctly predict a woman professor to be a professor (WPP).}
    \end{minipage}
\end{figure*}

\section{Selection of bios}\label{sec:selection_of_scenarios}

\paragraph{Pre-selection}
As outlined in \Cref{sec:study_setup}, participants are confronted with 14 bios of professors and teachers.
We impose a series of constraints to select which bios from the holdout set we include in the questionnaires.
In particular, for a given bio to be included in our questionnaires, we require it to satisfy the following:
\begin{itemize}
    \item Both models $\mathbf{AI}_{rel}$ and $\mathbf{AI}_{gen}$ must yield the same predicted occupation for the bio.
    \item The prediction probabilities of $\mathbf{AI}_{rel}$ and $\mathbf{AI}_{gen}$ towards either occupation must be \emph{at most} 20\% different. This ensures that both models are comparably certain in their predictions for the given bio.
    \item The prediction probabilities of $\mathbf{AI}_{rel}$ and $\mathbf{AI}_{gen}$ towards either occupation must be \emph{at most} 80\%. This aims at eliminating a large share of bios that are ``too easy'' to classify.
    \item To avoid any confounding effects of bios' length on people's behavior, we only consider bios of length between 50 and 100 words.
\end{itemize}
Enforcing these constraints on bios from the holdout set leaves us with 690 eligible bios (out of 16,221).
In a next step, we decide on the final set for our questionnaires.

\paragraph{Final selection}
The authors jointly screened these 690 bios and ruled out those that are trivial (e.g., because humans would easily be able to tell the occupation) or otherwise not suitable (e.g., because of misspellings or excessive use of jargon).
We also discarded bios where explanations would highlight too few or too many words, or where the number of highlighted words was significantly different between the \emph{task-relevant} and the \emph{gendered} condition.
This filtering narrows down the set of eligible bios to 38.
The authors then independently screened the resulting 38 bios including the corresponding explanations, and assigned a rating of green (``in favor of using it''), yellow (``indifferent''), or red (``in favor of discarding it''), based on both a bio's content as well as the associated explanation, favoring bios that were non-trivial but that contained enough information to make a correct prediction.
We then decided on the final set of 14 bios based on majority vote, taking into account the required composition of scenarios, as outlined in \Cref{tab:scenarios} in \Cref{sec:study_setup}.

\section{Overrides of anti-stereotypical AI recommendations}\label{sec:overrides_mtt_wpp}

\Cref{fig:mtt,fig:wpp} show participants' overriding behavior for cases where AI recommendations are correct and anti-stereotypical; i.e., correctly suggesting men to be teachers (MTT) and women to be professors (WPP).
We see that across conditions, overrides are much higher for the MTT case than for the WPP case.
Together with the findings from \Cref{sec:results} this suggests that participants were more prone to override AI recommendations whenever they suggested someone to be a teacher vs. a professor.

\end{document}